\documentclass[12pt]{article}
\input amssym.def
\input amssym.tex

\newcommand{\qed}{\rule{3mm}{3mm}}

\newcommand{\cF}{{\cal F}}

\newcommand{\rA}{{\rm A}}
\newcommand{\rF}{{\rm F}}
\newcommand{\rH}{{\rm H}}
\newcommand{\rL}{{\rm L}}
\newcommand{\rX}{{\rm X}}
\newcommand{\rY}{{\rm Y}}

\newcommand{\bL}{{\bf L}}
\newcommand{\bS}{{\bf S}}

\newcommand{\wip}{\widetilde{p}}
\newcommand{\wx}{\widetilde{x}}

\newcommand{\bbL}{{\Bbb L}}
\newcommand{\bbR}{{\Bbb R}}
\newcommand{\bbS}{{\Bbb S}}

\newtheorem{theorem}{Theorem}

\newbox\meibox
\def\placeunder#1#2#3#4{\setbox\meibox%
\vbox{\hbox{\hskip#4$\hphantom{#2}$}\hbox{$\hphantom{#1}$}}%
\vtop{\baselineskip=0pt\lineskiplimit=\baselineskip%
\lineskip=#3\hbox to \wd\meibox{\hfil\hskip#4$#2$\hfil}%
\hbox to \wd\meibox{\hfil$#1$\hfil}}}
\def\undertilde#1{\mathchoice{%
\placeunder{\vbox to 1.4pt{\hbox{$\displaystyle\widetilde{\,\,\,
}$}\vss}}{\displaystyle#1}{1.5pt}{1.5pt}}%
{\placeunder{\vbox to 1.4pt{\hbox{$\textstyle\widetilde{\,\,
}$}\vss}}{\textstyle#1}{1.5pt}{1.5pt}}%
{\placeunder{\vbox to 1.4pt{\hbox{$\scriptstyle\tilde{
}$}\vss}}{\scriptstyle#1}{1pt}{1pt}}%
{\placeunder{\vbox to 1.4pt{\hbox{$\scriptscriptstyle\tilde{
}$}\vss}}{\scriptscriptstyle#1}{1pt}{1pt}}%
}

\begin{document}

\begin{center}
{\large\bf 
Integrability of V.\,Adler's discretization of the Neumann system}
\end{center}
\vspace{0.5cm}

\begin{center}
{\sc Yuri B.\,Suris}\footnote{E--mail: 
{\tt suris} {\makeatother @ \makeatletter}{\tt sfb288.math.tu-berlin.de }}
\end{center}
\begin{center}
Fachbereich Mathematik, Technische Universit\"at Berlin, \\
Str. 17 Juni 136, 10623 Berlin, Germany
\end{center}
\vspace{0.5cm}

\begin{abstract}
We prove the integrability of the discretization of the Neumann system
recently proposed by V. Adler.
\end{abstract}

\newpage

\section{Introduction}\label{Sect Neumann introduction}

The famous Neumann system belongs to the area of constrained mechanics.
It describes the motion of a particle on the surface of the sphere 
$S=\{x\in{\Bbb R}^N:\,\langle x,x\rangle=1\}$ under the influence of the 
harmonic potential $\frac{1}{2}\langle \Omega x,x\rangle$, where
$\Omega={\rm diag}(\omega_1,\ldots,\omega_N)$. (Here and below
$\langle\cdot,\cdot\rangle$ stands for the standard Euclidean scalar product
in $\bbR^N$). The equations of motion read:
\begin{equation}\label{Neu introd}
\ddot{x}_k=-\omega_kx_k-\alpha x_k\;, \qquad 1\le k\le N\;,
\end{equation}
where $\alpha=\alpha(x,\dot{x})$ is the Lagrange multiplier assuring the 
validity of the relations $\langle x,x\rangle=1$, $\langle \dot{x},x\rangle=0$
during the evolution. It is easy to see that
\begin{equation}\label{Neu alpha introd}
\alpha=\langle \dot{x},\dot{x}\rangle-\langle \Omega x,x\rangle\;.
\end{equation}
This system can be given a Hamiltonian interpretation in terms of the Dirac
Poisson bracket, and it turns out to be completely integrable in the 
Liouville--Arnold sense, all integrals being quadratic in momenta (which also
implies that it is solvable via the separation of variables method). 
The literature on the Neumann system includes the original \cite{N}, as well
as modern presentations, e.g. \cite{Mo, R, F, Ma}.

The present paper is devoted to a discretization of (\ref{Neu introd}),
introduced recently by V. Adler in a beautiful paper \cite{A}, along with a 
novel discretization of the Landau--Lifschitz system. In difference equations
for $x:h{\Bbb Z}\mapsto S$, we write $x$ for $x(nh)$, $\wx$ for $x((n+1)h)$, 
and $\undertilde{x}$ for $x((n-1)h)$. The equations of motion of the V.Adler's
discretization read:
\begin{equation}\label{dNeu Adler introd}
\frac{\wx_k+x_k}{1+\langle\wx,x\rangle}+\frac{x_k+\undertilde{x}_k}
{1+\langle x,\undertilde{x}\rangle}=\frac{(2-\frac{h^2}{2}\omega_k)x_k}
{1-\frac{h^2}{4}\langle\Omega x,x\rangle}\;,\quad 1\le k\le N\;.
\end{equation}
While the integrability of this system for $N=3$ follows from Adler's results, 
the general case was left open by him. We construct explicitly the full set of 
integrals of motion for (\ref{dNeu Adler introd}).

It has to be said that several different integrable discretization of the 
Neumann system are currently known. The first one of them, due to A. Veselov 
\cite{V, MV} is governed by the following difference equations:
\begin{equation}\label{Neu Back introd}
\wx_k+\undertilde{x}_k=(1+h^2\omega_k)^{-1/2}\beta x_k\;,\quad 1\le k\le N\;,
\end{equation}
where $\beta$ is the Lagrange multiplier, assuring that $x$ remains on
the sphere $S$ during the discrete time evolution. It is easy to see 
that
\begin{equation}\label{Neu Back beta introd}
\beta=\frac{2\Big\langle (I+h^2\Omega)^{-1/2}x,\undertilde{x}\Big\rangle}
{\Big\langle (I+h^2\Omega)^{-1}x,x\Big\rangle}=
\frac{2\Big\langle (I+h^2\Omega)^{-1/2}x,\wx\Big\rangle}
{\Big\langle (I+h^2\Omega)^{-1}x,x\Big\rangle}=
\frac{\Big\langle (I+h^2\Omega)^{-1/2}x,\wx+\undertilde{x}\Big\rangle}
{\Big\langle (I+h^2\Omega)^{-1}x,x\Big\rangle}\;.
\end{equation}
The first two expressions for $\beta$ make this difference equation {\it
explicit} for the evolution in both the positive and the negative directions 
of (discrete) time. The third expression makes it obvious that
$\beta=2-h^2\alpha+o(h^2)$ with $\alpha$ from (\ref{Neu alpha introd}), so that
we indeed have a discretization of (\ref{Neu introd}). This
equation (\ref{Neu Back introd}) is a discrete time Lagrangian 
system, which allows one to introduce the canonically conjugate momenta
in a systematic way. Performing this, the main feature 
of (\ref{Neu Back introd}) comes into the light: namely, it shares the 
integrals of motion with its continuous time counterpart. In other words,
it is a B\"acklund transformation for the continuous time Neumann
system (cf. \cite{HKR}).

The second integrable discretization of the Neumann system, discovered by 
O. Ragnisco \cite{Rag}, is governed by the following difference equations:
\begin{equation}\label{dNeu New introd}
\frac{\wx_k}{\langle\wx,x\rangle}-2x_k+\frac{\undertilde{x}_k}
{\langle x,\undertilde{x}\rangle}=-h^2\omega_kx_k+h^2\langle\Omega x,x\rangle
x_k\;, \quad 1\le k\le N\;.
\end{equation}
Again, this is a discrete time Lagrangian system. Introducing the canonically
conjugate momenta in a proper way, one sees that the integrals of
motion for (\ref{dNeu New introd}) are $O(h)$--perturbations of the 
integrals of the continuous time Neumann system.

The comparison of the previously known discretizations was performed in
\cite{RS}. With the advent of the V. Adler's discretization and the proof of 
its integrability, the Neumann system becomes the best candidate for the 
championship in possessing various different integrable discretizations.

\setcounter{equation}{0}
\section{Neumann system}
\label{Sect Neumann}

The Neumann system describes the motion of a point $x\in{\Bbb R}^{N}$ 
under the potential $\frac{1}{2}\langle \Omega x,x\rangle$, constrained to the 
sphere 
\begin{equation}\label{Neu Sphere}
S=\Big\{x\in{\Bbb R}^{N}:\;\langle x,x\rangle=1\Big\}\,.
\end{equation}

The Lagrangian approach to this problem is as follows. The motions
should deliver local extrema to the action functional
\[
\bS=\int_{t_0}^{t_1}\bL\Big(x(t),\dot{x}(t)\Big)dt\;,
\]
where $\bL:TS\mapsto {\Bbb R}$ is the Lagrange function, given by
\begin{equation}
\bL(x,\dot{x})=\frac{1}{2}\langle\dot{x},\dot{x}\rangle-
\frac{1}{2}\langle \Omega x,x\rangle-\frac{1}{2}\,\alpha(\langle x,x\rangle
-1)\;.
\end{equation}
Here the first two terms on the right--hand side represent the unconstrained
Lagrange function, and the Lagrange multiplier $\alpha$ has to be chosen to
assure that the solution of the variational problem lies on the constrained
manifold $TS$, which is described by the equations
\begin{equation}\label{Neu Lagr constraints}
\langle x,x\rangle=1\;,\qquad \langle \dot{x},x\rangle=0\;.
\end{equation}
The differential equations of the extremals of the above problem read:
\begin{equation}\label{Neu}
\ddot{x}_k=-\omega_kx_k-\alpha x_k\;,
\end{equation}
or, in the vector form,
\begin{equation}\label{Neu vect}
\ddot{x}=-\Omega x-\alpha x\;.
\end{equation}
The value of $\alpha$ is determined as follows:
\[
0=\langle \dot{x},x\rangle^{\mbox{\boldmath$\cdot$}}=
\langle\dot{x},\dot{x}\rangle+\langle \ddot{x},x\rangle=
\langle\dot{x},\dot{x}\rangle-\langle\Omega x,x\rangle-\alpha\;.
\]
Therefore
\begin{equation}\label{Neu alpha}
\alpha=\langle \dot{x},\dot{x}\rangle-\langle\Omega x,x\rangle\;.
\end{equation}
So, the complete description of the Neumann problem is delivered by the
equations of motion (\ref{Neu}), augmented with the expression 
(\ref{Neu alpha}).

The Legendre transformation, leading to the Hamiltonian interpretation of 
the above system, is given by:
\[
\rH(x,p)=\langle \dot{x},p\rangle-\bL(x,\dot{x})\;,
\]
where the canonically conjugate momenta $p$ are given by
\begin{equation}\label{Neu p}
p=\partial\rL/\partial\dot{x}=\dot{x}\;.
\end{equation}
Hence we obtain
\begin{equation}\label{Neu H}
\rH(x,p)=\frac{1}{2}\langle p,p\rangle+
\frac{1}{2}\langle\Omega x,x\rangle\;.
\end{equation}
The corresponding symplectic structure is the restriction of the standard
symplectic structure of the space ${\Bbb R}^{2N}(x,p)$,
\begin{equation}\label{Neu PB unconstr}
\{p_k,x_j\}=\delta_{kj}\;,
\end{equation}
to the submanifold $T^*S$ which is singled out by the relations
\begin{equation}\label{Neu Ham constraints}
\phi_1=\langle x,x\rangle-1=0\;,\qquad \phi_2=\langle p,x\rangle=0\;.
\end{equation}
The Dirac Poisson bracket for this symplectic structure on $T^*S$ is 
given by the following relations:
\begin{equation}\label{Neu PB}
\{x_k,x_j\}_{\rm D}=0\;,\quad
\{p_k,x_j\}_{\rm D}=\delta_{kj}-\frac{x_kx_j}{\langle x,x\rangle}\;,
\quad
\{p_k,p_j\}_{\rm D}=\frac{x_kp_j-p_kx_j}{\langle x,x\rangle}\;.
\end{equation}
It is easy to check that the Hamiltonian vector field generated by the
Hamilton function (\ref{Neu H}) in the bracket $\{\cdot,\cdot\}_{\rm D}$
on $T^*S$, is given by:
\begin{equation}\label{Neu Ham}
\dot{x}_k=\{\rH,x_k\}_{\rm D}=p_k\;,\qquad
\dot{p}_k=\{\rH,p_k\}_{\rm D}=-\omega_kx_k-\alpha x_k\;,
\end{equation}
where
\begin{equation}\label{Neu alpha Ham}
\alpha=\langle p,p\rangle-\langle \Omega x,x\rangle\;.
\end{equation}
Obviously, this is nothing but the first--order form of (\ref{Neu}) with the 
multiplier (\ref{Neu alpha}).

Supposing that all $\omega_k$ are distinct (which will be assumed from now on), 
one can prove that the following $N$ functions are integrals of motion of 
(\ref{Neu Ham}): 
\begin{equation}\label{Neu Fk}
\rF_k=x_k^2+\sum_{j\neq k}\frac{(p_kx_j-x_kp_j)^2}
{\omega_k-\omega_j}\;,\qquad 1\le k\le N\;.
\end{equation} 
Only $N-1$ of them are functionally independent on $T^*S$, due to the relation
\begin{equation}\label{Neu int dep}
\sum_{k=1}^N \rF_k=\langle x,x\rangle\;,
\end{equation}
and this is equal to 1 on $T^*S$. The Hamilton function of the Neumann system
may be represented as
\begin{equation}\label{Neu int H}
\rH=\frac{1}{2}\sum_{k=1}^N \omega_k\rF_k
=\frac{1}{2}\Big(\langle p,p\rangle\langle x,x\rangle-
\langle p,x\rangle^2\Big)+\frac{1}{2}\langle \Omega x,x\rangle\;,
\end{equation}
which coincides with (\ref{Neu H}) on $T^*S$. All functions
$\rF_k$ are in involution, assuring the complete integrability
of the Neumann system with respect to the Dirac Poisson bracket
$\{\cdot,\cdot\}_{\rm D}$.

\section{Adler's discretization}
\label{Sect dNeumann Adler}

It will be convenient to denote
\begin{equation}
\alpha_k=1-\frac{h^2}{4}\,\omega_k\;,\qquad \rA={\rm diag}(\alpha_1,\ldots,
\alpha_N)=I-\frac{h^2}{4}\,\Omega\;.
\end{equation}
In these notations, the Adler's discretization (\ref{dNeu Adler introd}) is
written as
\begin{equation}\label{dNeu Adler}
\frac{\wx_k+x_k}{1+\langle\wx,x\rangle}+\frac{x_k+\undertilde{x}_k}
{1+\langle x,\undertilde{x}\rangle}=\frac{2\alpha_kx_k}
{\langle\rA x,x\rangle}\;.
\end{equation}
From the first glance, these equations look {\it implicit} for the evolution
in both the positive and the negative directions of time. However, this is
actually not the case. Consider, for definiteness, the case of the positive
time direction. We can express from (\ref{dNeu Adler}) the quantity
$1+\langle \wx,x\rangle$ through $x$ and $\undertilde{x}$. Indeed, rewrite
(\ref{dNeu Adler}) as 
\begin{equation}\label{dNeu Adler oncemore}
\frac{\wx+x}{1+\langle\wx,x\rangle}=
\frac{2\rA x}{\langle\rA x,x\rangle}-
\frac{x+\undertilde{x}}{1+\langle x,\undertilde{x}\rangle}\;.
\end{equation}
Taking the square of the norm of both sides, we find:
\[
\frac{2}{1+\langle\wx,x\rangle}=\left\|\frac{2\rA x}{\langle\rA x,x\rangle}-
\frac{x+\undertilde{x}}{1+\langle x,\undertilde{x}\rangle}\right\|^2\;,
\]
With this at hand, we can use (\ref{dNeu Adler oncemore}) to
express also $\wx$ as function of $x$ and $\undertilde{x}$.

By the way, the previous argument has an important corollary. It is easy
to see that the last formula may be represented as
\[
\frac{1}{1+\langle\wx,x\rangle}=-\frac{1}{1+\langle x,\undertilde{x}\rangle}+
\frac{2\langle\rA x,\rA x\rangle}{\langle\rA x,x\rangle^2}-
\frac{2\langle\rA x,\undertilde{x}\rangle}{\langle\rA x,x\rangle\,
(1+\langle x,\undertilde{x}\rangle)}\;.
\] 
However, we could as well perform a similar calculation for the evolution in
the negative direction of time, which would lead to the following formula:
\[
\frac{1}{1+\langle x,\undertilde{x}\rangle}=-\frac{1}{1+\langle\wx,x\rangle}+
\frac{2\langle\rA x,\rA x\rangle}{\langle\rA x,x\rangle^2}-
\frac{2\langle\rA x,\wx\rangle}{\langle\rA x,x\rangle\,
(1+\langle \wx,x\rangle)}\;.
\]
Comparing the last two formulas, we come to the following conclusion:
\begin{equation}
\frac{\langle\rA \wx,x\rangle}{1+\langle \wx,x\rangle}=
\frac{\langle\rA x,\undertilde{x}\rangle}{1+\langle x,\undertilde{x}\rangle}\;.
\end{equation}
In other words, we found an {\it integral of motion} of the difference
equation (\ref{dNeu Adler}):
\begin{equation}\label{dNeu Adler I1}
I_1(\wx,x)=\frac{2\langle\rA \wx,x\rangle}{1+\langle \wx,x\rangle}\;.
\end{equation}
The second integral of motion can be obtained even more straightforwardly.
Indeed, from (\ref{dNeu Adler oncemore}) there follows easily:
\begin{equation}
\frac{\langle\rA^{-1}(\wx+x),\wx+x\rangle}{(1+\langle\wx,x\rangle)^2}=
\frac{\langle\rA^{-1}(x+\undertilde{x}),x+\undertilde{x}\rangle}
{(1+\langle x,\undertilde{x}\rangle)^2}\;.
\end{equation}
In other words, the following function is also an {\it integral of motion} 
of the difference equation (\ref{dNeu Adler}):
\begin{equation}\label{dNeu Adler I2}
I_2(\wx,x)=
\frac{\langle\rA^{-1}(\wx+x),\wx+x\rangle}{(1+\langle\wx,x\rangle)^2}\;.
\end{equation}

\section{Lagrangian formulation}
\label{Sect dNeu Adler lagrangian}

We now demonstrate that (\ref{dNeu Adler}) may be interpreted as the equation
of extremals of a discrete time action functional
\[
\bbS=\sum_{n=n_0}^{n_1}\bbL\Big(x((n+1)h),x(nh)\Big)\;,
\]
where $\bbL:S\times S\mapsto{\Bbb R}$ is a discrete time Lagrange function.
Recall that the discrete time Lagrangian equations of motion read:
\begin{equation}\label{dLagr}
\nabla_x\bbL(\wx,x)+\nabla_x\bbL(x,\undertilde{x})=0\;,
\end{equation}
and that the momenta $p\in T^*_xS$, $\wip\in T^*_{\wx}S$ canonically conjugate 
to $x$, resp. to $\wx$, are given by:
\begin{eqnarray}
hp & = & -\nabla_x\bbL(\wx,x)\;,
\label{dLagr: p}\\ 
%\nonumber\\
h\wip & = & \nabla_{\wx}\bbL(\wx,x)\;.
\label{dLagr: wp}
\end{eqnarray}

Consider the discrete time Lagrange function on $S\times S$:
\begin{equation}\label{dNeu Lagr}
h\bbL(\wx,x)=-2\,\log(1+\langle\wx,x\rangle)+2\,\log\langle\rA x,x\rangle
\;,\quad x\in S,\; \wx\in S\,.
\end{equation}
Applying (\ref{dLagr: p}), (\ref{dLagr: wp}), we find:
\begin{eqnarray}
hp_k & = & \frac{2\wx_k}{1+\langle\wx,x\rangle}-\frac{4\alpha_kx_k}{\langle
\rA x,x\rangle}-\gamma x_k\;,  \label{dNeu Adler: p prelim}\\ 
%\nonumber\\
h\wip_k & = & -\frac{2x_k}{1+\langle\wx,x\rangle}+\delta\wx_k \;.
\label{dNeu Adler: wp prelim}
\end{eqnarray}
Here the scalar multipliers $\gamma$, $\delta$ have to be chosen so as to 
assure that
\[
p\in T_x^*S\;, \qquad \wip\in T_{\wx}^*S\;,
\]
or, in other words, to assure that the following relations hold:
\begin{equation}\label{dNeu cotang cond}
\langle p,x\rangle=0\;,\qquad \langle \wip,\wx\rangle=0\;.
\end{equation}
It is easy to see that this is achieved if
\begin{equation}\label{dNeu gamma}
\gamma=\frac{2\langle\wx,x\rangle}{1+\langle\wx,x\rangle}-4=
-\frac{2}{1+\langle\wx,x\rangle}-2\;,\qquad
\delta=\frac{2\langle\wx,x\rangle}{1+\langle\wx,x\rangle}=
-\frac{2}{1+\langle\wx,x\rangle}+2\;.
\end{equation}
So, the following are the Lagrangian equations of motion of the Adler's 
discretization of the Neumann system:
\begin{eqnarray}
hp_k & = & \frac{2(\wx_k+x_k)}{1+\langle\wx,x\rangle}-
\frac{4\alpha_kx_k}{\langle\rA x,x\rangle}+2x_k\;,
\label{dNeu Adler: p}\\ 
\nonumber\\
h\wip_k & = & -\frac{2(\wx_k+x_k)}{1+\langle\wx,x\rangle}+2\wx_k \;.
\label{dNeu Adler: wp}
\end{eqnarray}
Clearly, these two equations yield also the Newtonian form (\ref{dNeu Adler})
of the equations of motion.

The equations (\ref{dNeu Adler: p}), (\ref{dNeu Adler: wp}) define a 
symplectic map $(x,p)\in T^*S\mapsto(\wx,\wip)\in T^*S$ {\it explicitly}. 
Indeed, the first equation (\ref{dNeu Adler: p}) yields
\[
\frac{2(\wx_k+x_k)}{1+\langle\wx,x\rangle}=
\frac{4\alpha_kx_k}{\langle\rA x,x\rangle}-2x_k+hp_k\;,
\]
which implies
\begin{equation}\label{dNeu Adler norm}
\frac{2}{1+\langle\wx,x\rangle}=\left\|\frac{2\rA x}{\langle\rA x,x\rangle}-
x+\frac{hp}{2}\right\|^2\;.
\end{equation}
This, substituted back into (\ref{dNeu Adler: p}), allows us to 
determine $\wx$, and then, finally, the second equation of motion 
(\ref{dNeu Adler: wp}) defines $\wip$. (Clearly, this is a perifrase
of the similar argument in the previous section).

\section{Integrability}
\label{Sect dNeumann Adler integrability}

We can now express the integrals of motion $I_1$, $I_2$ in terms of canonically
conjugate variables $(x,p)$. Straightforward calculations verify the following
formulas:
\begin{equation}\label{dNeu Adler I1 x,p}
I_1(\wx,x)=\frac{2\langle \rA\wx,x\rangle}{1+\langle \wx,x\rangle}=
\langle \rA x,x\rangle-h\langle \rA x,p\rangle-\frac{h^2}{4}
\langle p,p\rangle\langle \rA x,x\rangle\;,
\end{equation}
\begin{equation}\label{dNeu Adler I2 x,p}
I_2(\wx,x)=\frac{\langle \rA^{-1}(\wx+x),\wx+x\rangle}{(1+\langle
\wx,x\rangle)^2}=
\langle \rA^{-1} x,x\rangle-h\langle \rA^{-1} x,p\rangle+\frac{h^2}{4}
\langle \rA^{-1}p,p\rangle\;.
\end{equation}
These two integrals are enough to assure the Liouville--Arnold integrability
of the map (\ref{dNeu Adler: p}), (\ref{dNeu Adler: wp}) for $N=3$. The
full set of integrals in the general case is given in the following statement,
which constitutes the main result of this Letter.

\begin{theorem}\label{dNeumann Adler integrals}
If all $\alpha_k$'s are distinct, then the following functions are integrals
of motion of the Adler's discrete time Neumann system:
\begin{equation}\label{dNeu Adler Fk}
\cF_k=x_k^2-hx_kp_k+\frac{h^2}{4}\sum_{j\neq k}
\frac{(x_kp_j-x_jp_k)(\alpha_kx_kp_j-\alpha_jx_jp_k)}{\alpha_j-\alpha_k}\;.
\end{equation}
\end{theorem}
{\bf Proof.} Denote
\begin{equation}\label{dNeu Adler XY}
\rX_{kj}=x_kp_j-x_jp_k\;, \qquad \rY_{kj}=\alpha_kx_kp_j-\alpha_jx_jp_k\;.
\end{equation}
We have:
\begin{equation}\label{dNeu Adler aux1}
\sum_{j\neq k}\frac{\widetilde{\rX}_{kj}\widetilde{\rY}_{kj}}
{\alpha_j-\alpha_k}
-\sum_{j\neq k}\frac{\rX_{kj}\rY_{kj}}{\alpha_j-\alpha_k}=
\sum_{j\neq k}\frac{\widetilde{\rX}_{kj}(\widetilde{\rY}_{kj}-
\rY_{kj})+\rY_{kj}(\widetilde{\rX}_{kj}-\rX_{kj})}{\alpha_j-\alpha_k}\;.
\end{equation}
Using the equations of motion (\ref{dNeu Adler: p}), (\ref{dNeu Adler: wp}),
we find:
\begin{eqnarray*}
\frac{h}{2}\,\widetilde{\rX}_{ij} & = & \frac{x_k\wx_j-\wx_kx_j}
{1+\langle\wx,x\rangle}\;,\\ \\
\frac{h}{2}\,\rY_{kj} & = & \frac{\alpha_kx_k\wx_j-\alpha_j\wx_kx_j+
(\alpha_k-\alpha_j)x_kx_j}{1+\langle\wx,x\rangle}+(\alpha_k-\alpha_j)x_kx_j\;, 
\end{eqnarray*}
and
\begin{eqnarray*}
\frac{h}{2}\,\frac{\widetilde{\rX}_{kj}-\rX_{kj}}{\alpha_j-\alpha_k} & = & 
\frac{2x_kx_j}{\langle\rA x,x\rangle}\;,
\\ \\
\frac{h}{2}\,\frac{\widetilde{\rY}_{kj}-\rY_{kj}}{\alpha_j-\alpha_k} & = & 
\frac{x_kx_j+\wx_kx_j+x_k\wx_j+\wx_k\wx_j}
{1+\langle\wx,x\rangle}+x_kx_j-\wx_k\wx_j\;.
\end{eqnarray*}
Calculating further, we find:
\begin{eqnarray*}
\lefteqn{\frac{h^2}{4}\,\sum_{j\neq k}\frac{\widetilde{\rX}_{kj}
(\widetilde{\rY}_{kj}-\rY_{kj})}{\alpha_j-\alpha_k}=}\\
& = & \frac{1}{1+\langle\wx,x\rangle}\sum_{j=1}^N
(x_k\wx_j-\wx_kx_j)\bigg(\frac{x_kx_j+\wx_kx_j+x_k\wx_j+\wx_k\wx_j}
{1+\langle\wx,x\rangle}+x_kx_j-\wx_k\wx_j\bigg)\\
& = & \frac{1}{1+\langle\wx,x\rangle}\Big(x_k^2(1+\langle\wx,x\rangle)-
2\wx_kx_k-\wx_k^2(1-\langle\wx,x\rangle)\Big)\\
& = & x_k^2-\wx_k^2+h\wx_k\wip_k\;;
\end{eqnarray*}
\begin{eqnarray*}
\lefteqn{\frac{h^2}{4}\,\sum_{j\neq k}\frac{\rY_{kj}
(\widetilde{\rX}_{kj}-\rX_{kj})}{\alpha_j-\alpha_k}=}\\
& = & \frac{2}{\langle\rA x,x\rangle}\sum_{j=1}^N x_kx_j
\bigg(\frac{\alpha_kx_k\wx_j-\alpha_j\wx_kx_j+
(\alpha_k-\alpha_j)x_kx_j}{1+\langle\wx,x\rangle}+(\alpha_k-\alpha_j)x_kx_j
\bigg)\\
& = & \frac{2x_k}{\langle\rA x,x\rangle}\bigg(2\alpha_kx_k-
\frac{(\wx_k+x_k)}{1+\langle\wx,x\rangle}\,\langle\rA x,x\rangle
-x_k\langle\rA x,x\rangle\bigg)\\
& = & -hx_kp_k\;.
\end{eqnarray*}
Collecting all results, we find:
\[
\frac{h^2}{4}\sum_{j\neq k}\frac{\widetilde{\rX}_{kj}\widetilde{\rY}_{kj}}
{\alpha_j-\alpha_k}
-\frac{h^2}{4}\sum_{j\neq k}\frac{\rX_{kj}\rY_{kj}}{\alpha_j-\alpha_k}
=x_k^2-\wx_k^2+h\wx_k\wip_k-hx_kp_k\;.
\]
This proves the Theorem. \qed

The involutivity of the integrals $\cF_k$ may be established by the
arguments similar to those in \cite{Ma}. It is easy to see that the two 
previously found integrals (\ref{dNeu Adler I1 x,p}), (\ref{dNeu Adler I2 x,p}) 
admit the following expression in terms of $\cF_k$:
\[
2I_1=\sum_{k=1}^N\alpha_k\cF_k\;,\qquad I_2=\sum_{k=1}^N \alpha_k^{-1}\cF_k\;.
\]

\section{Conclusions}

Several things remain to be done to put the V. Adler's discrete time
Neumann system into the modern framework of discrete integrable systems:
to find Lax representations and their $r$--matrix interpretation, and
to perform an integration in terms of theta--functions. I hope to have
an opprtunity to present these results elsewhere. I thank V. Adler and
A. Shabat for showing me their results prior to publication.

%%%%%%%%%%%%%%%%%%%%%%%%%%%%%%%%%%%%%%%%%%%%%%%%%%%%%%%%%%%%%%%%%%%%%%%%%%

\end{document}